\RequirePackage{fix-cm}
\documentclass{svjour3}                     
\smartqed  
\usepackage{mathtext}
\usepackage{mathenv}
\usepackage{float}
\usepackage{array}
\usepackage{graphicx, psfrag}
\usepackage{amsmath,amssymb}
\usepackage[left=3cm,right=3cm,top=2.0cm,bottom=1.5cm]{geometry}
\begin{document}

\title{Simulation of fiber-reinforced viscoelastic structures
subjected to finite strains: multiplicative approach
}
\subtitle{Simulation of fiber-reinforced viscoelastic structures }


\author{I.I. Tagiltsev   \and
        P.P. Laktionov  \and
        A.V. Shutov  
}


\institute{ I.I. Tagiltsev \at
              Lavrentyev Institute of Hydrodynamics, Pr. Lavrentyeva 15, 630090 Novosibirsk, Russia \\
              Novosibirsk State University, Pirogova 1, 630090 Novosibirsk, Russia \\
              Tel.: +7-383-333-17-46\\
              Fax: +7-383-333-16-12\\
              \email{gannitgl@gmail.com}
               \and
              P.P. Laktionov \at
              Institute of Chemical Biology and Fundamental Medicine, Siberian Branch, Russian Academy of Sciences, 630090 Novosibirsk, Russia \\
              Tel.: +7-383-363-51-43 \\
              Fax: +7-383-363-51-53 \\
              \email{lakt@niboch.nsc.ru}           
           \and
              A.V. Shutov \at
              Lavrentyev Institute of Hydrodynamics, Pr. Lavrentyeva 15, 630090 Novosibirsk, Russia \\
              Novosibirsk State University, Pirogova 1, 630090 Novosibirsk, Russia \\
              Tel.: +7-383-333-17-46\\
              Fax: +7-383-333-16-12\\
              \email{alexey.v.shutov@gmail.com}           
}

\date{Received: date / Accepted: date}

\maketitle

\begin{abstract}
The study is devoted to the
 geometrically nonlinear simulation of fiber-reinforced composite structures.
The applicability of the multiplicative approach to the simulation of viscoelastic properties
of a composite material is assessed, certain improvements are suggested.
For a greater accuracy in applications involving local compressive fiber buckling,
a new family of hyperelastic potentials is introduced.
This family allows us to account for the variable critical compressive stress, which depends on the fiber-matrix interaction.
For the simulation of viscoelasticity, the well-established Sidoroff decomposition of the deformation gradient is implemented.
To account for the viscosity of the matrix material, the model of Simo and Miehe (1992) is used; highly efficient
iteration-free algorithms are implemented. The viscosity of the fiber is likewise described by
the multiplicative decomposition of the deformation gradient, leading to a scalar differential equation;
an efficient iteration-free algorithm is proposed for the implicit time stepping.
The accuracy and convergence of the new iteration-free method is tested and compared to that of
the standard scheme implementing the Newton iteration.
To demonstrate the applicability of the approach, a pressurized
multi-layer composite pipe is modelled; the so-called
stretch inversion phenomenon is reproduced and explained.
The stress distribution is obtained by a semi-analytical procedure; it may serve as a benchmark for FEM computations.
Finally, the issue of the parameter identification is addressed.
\end{abstract}


\section*{Nomenclature}

\begin{table}[H]
\begin{tabular}{rl}
$\mathbf{F}$ & deformation gradient\\
$\mathbf{C}$ & right Cauchy-Green tensor\\
$\mathbf{L}$ & velocity gradient tensor\\
$\mathbf{D}$ & strain rate tensor\\
$\mathbf{1}$ & identity tensor\\
$\mathbf{\tilde{T}}$ & 2nd Piola-Kirchhoff stress tensor\\
$\mathbf{S}$ & Kirchhoff stress tensor\\
$\mathbf{T}$ & Cauchy stress tensor\\
$\psi$ & Helmholtz  free-energy per unit mass\\
$\text{tr}(\mathbf{A})$ & trace of a tensor\\
$\mathbf{A}^{\text{D}}$ & deviatoric part of a tensor\\
$\mathbf{\overline{A}}$ & unimodular part of a tensor \\
\end{tabular}
\end{table}

\section{Introduction}
\label{intro}

Fiber-reinforced composite materials are encountered in various applications;
materials which can sustain large cyclic strains are of particular interest.
Rubber matrix composites are utilized as components of tires and air suspension in the automotive
 industry.
Various connective tissues especially those localized at the borders of organs, tendons and different types of
cartilage \cite{Springhetti} can be seen
as fiber-reinforced composites as well. Dealing with them, stiff
collagen fibers are typically idealized by
one-dimensional fiber families, which may possess different orientations.
Both in engineering and bio-mechanics, a common feature of such materials is that
a relatively stiff fiber is submerged into a soft isotropic matrix thus grunting
the material a set of desired properties not available for monolithic materials.
Following this idea, a number of biocompatible scaffolds produced by electrospinning and
able to be populated by living cells are offered for vascular graft or heart valves production.
Reinforced by fibers and filled with extracellular matrix produced by cells they are considered
to possess the demanded mechanical and biological properties \cite{AntonioDAmore2018}.

Global strength analysis of such structures involves large-scale numerical simulations;
on the macroscopic level the composite materials are idealized
as a homogeneous anisotropic visco-elastic continuum. Typically,
the mechanical response of these materials is highly anisotropic and viscous.
Interaction between geometric and physical nonlinearities as well as evolving anisotropy
place heavy demands on the applied conceptual framework.
Therefore, the classic triad 'model-algorithm-identification' is analyzed and discussed here.

Due to the occurrence of large strains, geometric nonlinearities need to be taken into account properly.
Hyperelasticity is the only thermodynamically consistent way to model the elastic properties of the material.
For isotropic hyperelastic strain energy functions
in application to rubber-like materials and soft biological tissues
the reader is referred to \cite{Beatty1987}, \cite{Delfino1997}; for anisotropic strain
energy functions in application to reinforcing fibers see
 \cite{Fung1983}, \cite{Vaishnav1973}, \cite{Holzapfel2000}.
A recent review of hyperelastic potentials used in bio-mechanics is
presented in \cite{Chagnon}.
Under compressive stresses, the reinforcing fibers may buckle so it is important
to model this local instability phenomenon within an appropriate material model.
The critical compressive stress depends on the architecture of reinforcement as well as on the fiber/matrix interaction.
In the current study a family of strain energy functions is suggested with adjustable critical stress
to account for these effects more accurately on the macrostructural level.

For the geometrically exact simulation of the visco-elastic properties, some additional assumptions need to be made concerning
the decomposition of the total strain into elastic and viscous parts, energy storage, flow rules etc.,
\cite{Holzapfel2001}, \cite{Holzapfel2002},
 \cite{Waffenschmidt2014}, \cite{Nedjar2011}, \cite{Li2015}. In this work we
advocate a modelling approach based on the \emph{multiplicative decomposition of the deformation gradient tensor}.
As discussed in \cite{Shutov2014}, \cite{Shutov2016},
this approach has numerous advantages over competing alternatives.
In particular, it allows us to build material models which are thermodynamically consistent and objective. A pure
split of the stress response in the deviatoric and volumetric parts can be enforced;
it enables us to model incompressible behaviour in a straightforward way.
Another important aspect is the possibility of an efficient numerical integration of the underlying constitutive
equations. For the multiplicative Maxwell body with an isotropic strain energy function,
efficient algorithms were already reported in
\cite{Shutov2013}, \cite{Shutov2017}. In the current study, an efficient algorithm is discussed
 dealing with a certain class of anisotropic strain energy
functions.

A demonstration problem involving the inflation of a
pressurized composite pipe comprising multiple anisotropic layers is analyzed.
In engineering, such problems may arise in analysis of pneumatic suspension;
 in bio-mechanics it is related to the
behaviour of blood vessels or blood vessel prosthesis. Actually, different
compliance of natural arteries and vascular grafts is the main reason of stenosis and short-term
patency of vascular grafts \cite{Stewart1992}, \cite{Salacinski2001}.
The so-called stretch inversion phenomenon is an important aspect which may appear at the initial stage of the deformation of
fiber-reinforced tubes (cf. \cite{Wiesemann1995}). This phenomenon is explained and modelled by the composite model.

The paper outline is as follows. In Section 2, a geometrically
exact model of a fiber-reinforced composite material is formulated. Next, in Section 3,
we discuss its efficient numerical implementation. A demonstration problem concerning the inflation of
a pressurized composite tube is solved in Section 4
with a short insight into material parameter identification.
Finally, concluding remarks are given in Section 5.

\section{Composite material model}

\begin{figure}\centering
    \psfrag{PsiMR}[m][][1][0]{$\Psi_{MR}$}
    \psfrag{PsiFiber}[m][][1][0]{$\Psi_{fiber}$}
    \psfrag{PsiNH}[m][][1][0]{$\Psi_{neo-Hooke}$}
    \psfrag{PsiFiber2}[m][][1][0]{$\Psi_{fiber}^{visc}$}
    \scalebox{1.0}{\includegraphics{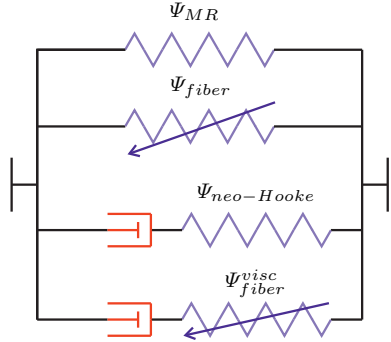}}
    \caption{An idealized rheological interpretation of the composite model based on the iso-strain approach: both the matrix and the fibers are
    subjected to the same strain. \label{ReologComposite}}
\end{figure}

In this work we focus on composite materials which exhibits visco-elastic properties.
Within an iso-strain approach, isotropic parts of the model (corresponding to the matrix material) are reinforced by
anisotropic parts (related to the viscous fibers). A vivid rheological interpretation
 of the iso-strain approach is provided by the concept of parallel connection shown in Figure \ref{ReologComposite}.

Let $\mathbf{F}$ be the deformation gradient which transforms line elements from the local reference
configuration $\tilde{\mathcal{K}}$ to the current configuration $\mathcal{K}$. We assume that the basic idealized
elements shown in Figure \ref{ReologComposite} are subjected to the same  deformation, described by $\mathbf{F}$.
The overall right Cauchy Green tensor is given by $\mathbf{C} := \mathbf{F}^{\text{T}} \mathbf{F}$. Its unimodular
part equals $\overline{\mathbf{C}} := (\det \mathbf{C})^{-1/3} \mathbf{C}$.

\subsection{Isotropic hyperelasticity}

In order to describe the isotropic part of the hyperelastic response, the well-known Mooney-Rivlin potential is used, which corresponds to
the following form of the Helmholz free energy per unit mass
 \begin{equation}\label{MooneyRivlin}
    \ \Psi_{MR}(\overline{\mathbf{C}}) = \frac{\displaystyle c_{1}}{2}(\overline{I}_{1} -3) + \frac{\displaystyle c_{2}}{2}(\overline{I}_{2} - 3).
\end{equation}
Here, $c_{1}$ are $c_{2}$ are the shear moduli of the matrix material, $\overline{I}_{1} := \overline{\mathbf{C}}:\mathbf{1}$ and
$\overline{I}_{2} := \frac{\displaystyle 1}{\displaystyle 2} ((\overline{\mathbf{C}}:\mathbf{1})^{2} - \mathbf{\overline{C}}^{2}:\mathbf{1})
 = \overline{\mathbf{C}}^{-1}:\mathbf{1}$. The model reduces to the
classical neo-Hookean material as $c_{2} \rightarrow 0$. The Mooney-Rivlin potential is frequently used to model
rubber-like materials and biological soft tissues (see, for example \cite{Koolstra}).

\subsection{Hyperelasticity of fibers}
Now let us consider a family of fibers which are directed in the reference configuration along the unit vector $\tilde{\mathbf{a}}$.
Let $\lambda^{(\tilde{\mathbf{a}})}$ be
the stretch of these fibers. It is computed as follows:
\begin{equation}
    \lambda^{(\tilde{\mathbf{a}})} := \| \mathbf{F} \tilde{\mathbf{a}} \|, \quad \Rightarrow \quad (\lambda^{(\tilde{\mathbf{a}})})^{2} = \overline{\mathbf{C}}:( \tilde{\mathbf{a}} \otimes \tilde{\mathbf{a}}) = \tilde{\mathbf{a}} \cdot \overline{\mathbf{C}} \cdot \tilde{\mathbf{a}}.
\end{equation}
For hyperelastic fibers it is natural to assume that the Helmholz free energy per unit mass depends on their stretch
\begin{equation}
    \Psi_{fiber} = \Psi_{fiber}(\lambda^{2}).
\end{equation}
For the following it is convenient to introduce the derivative
 \begin{equation}
 f := \frac{\displaystyle d \ \Psi_{fiber}(\lambda ^{2}) }{\displaystyle d ( \lambda ^{2} )}.
 \end{equation}
One of the most common potentials was proposed for hyperelastic fibers by Holzapfel et al. in \cite{Holzapfel2000}
\begin{equation} \label{HolzapfelAniso}
    \begin{array}{l}
        \displaystyle \ \Psi_{Holzapfel}(\lambda ^{2}) = \frac{k_{1}}{2 k_{2}} (e^{k_{2}(\lambda^{2}-1)^{2}} - 1), \quad
        f = 2k_{1}(\lambda^{2} - 1)e^{ k_{2}(\lambda^{2} - 1)^{2} }.
    \end{array}
\end{equation}
Here, $k_{1} \geq 0$ is the stiffness parameter; $k_{2} > 0$ is a non-dimensional parameter attributed to the nonlinearity
of the stress response under uniaxial tension.
The potential meets the natural requirement that the material is stress free in undeformed state.
An important property is that stresses grow exponentially under tension and remain bounded under compression.

This one-dimensional model of fibers is intended for use within a more general composite model.
Dealing with fiber-reinforced materials, the fibers may buckle under certain compressive loads,
which depend on the fiber-matrix interaction and other structural parameters.
When working with representative volume elements, some of these structural
instabilities can be naturally reproduced (cf. \cite{Goldberg}).
However, within the homogeneous approach used here, the instabilities need
to be explicitly introduced on the local (material) level.
Toward that end we consider some generalizations of \eqref{HolzapfelAniso}.

\subsubsection{Potential with fiber slackness}
It is possible to use a modification of the Holzapfel potential with three parameters, similar to the one used by von Hoegen et al. в \cite{Hoegen2017}
\begin{equation} \label{PsiMarino}
\begin{array}{l}
    \Psi_{Holzapfel*} = \frac{k_{1}}{k_{2}} ( e^{ k_{2} \langle \lambda_{\text{e}}^{2} - k_{3} - 1 \rangle ^{2} } - 1), \quad
    f_{Holzapfel*} = 2k_{1} \langle \lambda_{\text{e}}^{2} - k_{3} - 1 \rangle e^{ k_{2}\langle \lambda_{\text{e}}^{2} - k_{3} - 1 \rangle^{2} }.
\end{array}
\end{equation}
Here, $\langle a \rangle := \max \{ a,0 \} $; $ k_{1}$ and $k_{2}$
have the same meaning as for the Holzapfel potential; $k_{3} \geq 0$ is a slackness parameter
since for any stretches below
$\sqrt{1+k_{3}}$ the stresses are zero.
Unfortunately, for this approach the function $\Psi_{Holzapfel*}$
and its derivative are not smooth. Moreover, this potential yields zero stresses under compression which
is not always the case.

The next approach represents an interpolation between \eqref{HolzapfelAniso} and \eqref{PsiMarino} in case $k_{3} = 0$.

\subsubsection{Arctan-potential}

Let us consider the following ansatz for the Helmholz free energy:
\begin{equation}
    f_{arctan}(\lambda_{\text{e}}^{2}) = 2k_{1} (\lambda_{\text{e}}^{2} - 1)
    e^{ k_{2} (\lambda_{\text{e}}^{2} - 1)^{2} } \frac{\arctan{k_{3}(\lambda_{\text{e}}^{2} - 1)} + \pi/2}{\pi}.
\end{equation}
Thus, we have
\begin{equation} \label{PsiArctan}
    \Psi_{arctan}(\lambda_{\text{e}}^{2}) = \int_{0}^{\lambda_{\text{e}}^{2}} f_{arctan}(\xi) d \xi.
\end{equation}
Here, the parameter $k_{3} \geq 0$ is used to adjust the critical compressive stress corresponding to the fiber buckling. The potential $\Psi_{arctan}$ tends to $\Psi_{Holzapfel*}$ given in \eqref{PsiMarino} as $k_{3} \rightarrow 0$.

Moreover, the critical compressive stress can be adjusted by combining one of the modified potentials \eqref{PsiMarino} or \eqref{PsiArctan} with the
original potential \eqref{HolzapfelAniso}:
\begin{equation}\label{GeneralizedPotential}
    \Psi_{fiber} = \alpha \Psi_{Holzapfel} + (1-\alpha) \Psi_{modif},
\end{equation}
where $ 0 \leq \alpha \leq 1$ is a non-dimensional weighting coefficient; $\Psi_{modif} \in \{ \Psi_{Holzapfel*}$, $\Psi_{arctan} \}$.

\subsection{Isotropic Maxwell body} \label{Ce}

In order to account for the viscous properties of the material we
implement a certain finite strain formulation of the Maxwell body, initially proposed by
Simo \& Miehe in \cite{SimoMiehe}. For this model, the kinematics is based on the Sidoroff assumption, which
implies the multiplicative split of the deformation gradient tensor $\mathbf{F}$ into the elastic part $\hat{\mathbf{F}}_{\text{e}}$
and the inelastic part $\mathbf{F}_{\text{\text{i}}}$
\begin{equation}
\mathbf{F} = \hat{\mathbf{F}}_{\text{e}} \mathbf{F}_{\text{i}}.
\end{equation}
Basing on this decomposition we obtain tensors of right Cauchy-Green
type (RCGT): the inelastic RCGT $ \mathbf{C}_{\text{i}} = \mathbf{F}^{\text{T}}_{\text{i}}\mathbf{F}_{\text{i}}$
and the elastic RCGT $ \hat{\mathbf{C}}_{\text{e}} = \hat{\mathbf{F}}^{\text{T}}_{\text{e}}\hat{\mathbf{F}}_{\text{e}}$.
We note that the elastic tensor $ \hat{\mathbf{C}}_{\text{e}} $ operates on the stress-free intermediate configuration
$\hat{\mathcal{K}} := \mathbf{F}_{\text{i}} \tilde{\mathcal{K}}$.
Next,
we introduce the inelastic velocity gradient:
$ \hat{\mathbf{L}}_{\text{i}} = \dot{\mathbf{F}}_{\text{i}}\mathbf{F}^{-1}_{\text{i}} $. Its symmetric part is referred to as
the inelastic strain rate: $ \hat{\mathbf{D}}_{\text{i}} = \frac{1}{2}(\hat{\mathbf{L}}_{\text{i}} + \hat{\mathbf{L}}_{\text{i}}^{\text{T}}) $.
Further, using the Cuachy stress (true stress) $\mathbf{T}$, we introduce the Kirchhoff stress $\mathbf{S}$,
the second Piola-Kirchhoff stress $\hat{\mathbf{S}}$ acting on the intermediate configuration
 $\hat{\mathcal{K}}$, and the second Piola-Kirchhoff tensor $ \tilde{\mathbf{T}}$ operating on the reference configuration:
 \begin{equation}
\mathbf{S} := \det(\mathbf{F})\mathbf{T},  \quad \hat{\mathbf{S}} := \hat{\mathbf{F}}^{-1}_{\text{e}}\mathbf{S}\hat{\mathbf{F}}^{-\text{T}}_{\text{e}}, \quad
        \tilde{\mathbf{T}} := \mathbf{F}^{-1}\mathbf{S}\mathbf{F}^{-\text{T}}.
\end{equation}

For the elastic properties we assume a neo-Hookean potential; using the Coleman-Noll procedure
we arrive at the following expression for the second Piola-Kirchhoff tensor:
\begin{equation} \label{IsotrNeoHook}
    \begin{array}{l}
        \displaystyle \rho_{R}\Psi_{neo-Hooke}(\hat{\mathbf{C}}_{\text{e}}) =
        \frac{\mu}{2}(\text{tr}  \overline{\hat{\mathbf{C}}_{\text{e}}} -3), \quad
        \displaystyle \hat{\mathbf{S}} = 2 \rho_{R} \frac{\partial \Psi_{neo-Hooke}(\hat{\mathbf{C}}_{\text{e}}) }{\partial \hat{\mathbf{C}}_{\text{e}}}.
    \end{array}
\end{equation}
Here $\rho_{R}$ stands for the mass density in the reference configuration; $\mu$ is the shear modulus.

Now we consider the Clausius-Duhem inequality. In the isothermic case it reads
 \begin{equation}\label{Duhem_ini}
 \delta_{\text{i}} = \frac{1}{2 \rho_{R}} \tilde{\mathbf{T}}:\dot{\mathbf{C}} - \dot{\Psi}_{neo-Hooke} \ge 0.
\end{equation}
Using the previous assumptions, after some tensor algebra it can be reduced to
 \begin{equation}\label{Duhem}
        \rho_{R}\delta_{\text{i}} = (\hat{\mathbf{C}}_{\text{e}}\hat{\mathbf{S}}):\hat{\mathbf{D}}_{\text{i}} \ge 0.
\end{equation}
Next, we employ the flow rule which identically satisfies inequality \eqref{Duhem}:
 \begin{equation}\label{FlowRule}
  \displaystyle \hat{\mathbf{D}}_{\text{i}} = \frac{1}{2\eta}(\hat{\mathbf{C}}_{\text{e}}\hat{\mathbf{S}})^{\text{D}}.
\end{equation}
It implies an incompressible flow: $ \text{tr}(\hat{\mathbf{D}}_{\text{i}}) = 0 $.
Further, we re-formulate the constitutive equations by pulling relations \eqref{IsotrNeoHook} and \eqref{FlowRule} back to the reference configuration:

 \begin{equation}\label{Cinelast}
    \begin{array}{l}
        \displaystyle \Psi_{neo-Hooke} = \Psi_{neo-Hooke}(\mathbf{C}\mathbf{C}^{-1}_{\text{i}}) = \frac{\mu}{2\rho_{R}} (\text{tr}(\overline{\mathbf{C}\mathbf{C}^{-1}_{\text{i}}}) -3),\\
        \displaystyle \tilde{\mathbf{T}} = 2 \rho_{R} \frac{\partial \Psi_{neo-Hooke}(\mathbf{C}\mathbf{C}^{-1}_{\text{\text{i}}}) }{\partial \mathbf{C}} \mid_{\mathbf{C}_{\text{i}} = const}, \quad
        \tilde{\mathbf{T}} = \mu \mathbf{C}^{-1} (\overline{\mathbf{C}}\mathbf{C}^{-1}_{\text{i}})^{\text{D}}, \\
        \displaystyle \dot{\mathbf{C}}_{\text{i}} = \frac{1}{\eta}(\mathbf{C}\tilde{\mathbf{T}})^{\text{D}}\mathbf{C}_{\text{i}} = \frac{\mu}{\eta}(\overline{\mathbf{C}}\mathbf{C}^{-1}_{\text{i}})^{\text{D}}\mathbf{C}_{\text{i}}.
    \end{array}
\end{equation}
The system of constitutive equations is closed by specifying initial conditions
 \begin{equation}\label{Cinelast2}
     \mathbf{C}_{\text{i}}\mid_{t = t^{0}} = \mathbf{C}^{0}_{\text{i}}.
\end{equation}
Due to the isotropy of the elastic properties, the inelastic spin $\text{skew}(\mathbf{L}_{\text{i}})$ remains undetermined.
Effectively, this allows us to build the flow rule in six dimensions only.

\subsection{Fiber-like Maxwell body}

The fourth constituent of the composite model is the part governing the viscosity of fibers.
We start from a similar multiplicative split $ \mathbf{F} = \hat{\mathbf{F}}_{\text{e}} \mathbf{F}_{\text{i}} $, where the parts
$\hat{\mathbf{F}}_{\text{e}}$ and $\mathbf{F}_{\text{i}}$ have a similar meaning as in the previous subsection.
Recall that the fiber direction in the reference configuration is given by the unit vector $\tilde{\mathbf{a}}$ ($\| \tilde{\mathbf{a}} \| = 1 $).
Naturally, the action of the deformation gradient on $\tilde{\mathbf{a}}$ defines the stretch of the fiber:
\begin{equation}
   \lambda :=  \| \mathbf{F} \tilde{\mathbf{a}} \|  > 0.
\end{equation}
Assume that the inelastic part $\mathbf{F}_{\text{i}}$ satisfies the following condition:
$ \mathbf{F}_{\text{i}} \tilde{\mathbf{a}} = \lambda_{\text{i}} \tilde{\mathbf{a}}$.
In other words we impose the restriction that $\tilde{\mathbf{a}}$ is an eigenvector of
 $\mathbf{F}_{\text{i}}$. The elastic stretch $\lambda_{\text{e}}$
 is determined through
 \begin{equation}\label{lambdaei}
 \lambda_{\text{e}} = \| \mathbf{F}_{\text{e}} \tilde{\mathbf{a}} \| .
 \end{equation}
Then we arrive at:
\begin{equation}
 \lambda = \| \mathbf{F} \tilde{\mathbf{a}} \| = \|\mathbf{F}_{\text{e}} \mathbf{F}_{\text{i}}
 \tilde{\mathbf{a}}\| = \lambda_{\text{i}} \|\mathbf{F}_{\text{e}} \tilde{\mathbf{a}}\| = \lambda_{\text{i}} \lambda_{\text{e}}.
\end{equation}
It is natural to assume that the free energy depends on the elastic stretch $\lambda_{\text{e}}$, which can be evaluated through
\begin{equation}
  \lambda^{2}_{\text{e}} = \overline{\mathbf{C}}_{\text{e}}:\mathbf{M},
\end{equation}
where $ \mathbf{M} := \tilde{\mathbf{a}} \otimes \tilde{\mathbf{a}} $.
Similar to the isotropic setting,
the Coleman-Noll procedure yields the following relation on
the stress-free (intermediate) configuration
\begin{equation}
 \hat{\mathbf{S}} = 2 \rho_{R} \frac{\displaystyle \partial \Psi_{fiber}^{visc} }{\displaystyle \partial \mathbf{C}_{\text{e}}}.
\end{equation}
Algebraic transformations yield
 \begin{equation}\label{fiberElas}
  \hat{\mathbf{S}} = 2 \rho_{R} f \mathbb{P}_{\mathbf{C}_{\text{e}}}:\mathbf{M},
 \end{equation}
where
 \begin{equation}\label{fiberElas2}
 f = \frac{\displaystyle \partial \Psi_{fiber}^{visc}(\overline{\mathbf{C}}_{\text{e}}:\mathbf{M}) }
 {\displaystyle \partial (\overline{\mathbf{C}}_{\text{e}}:\mathbf{M})}, \quad \mathbb{P}_{\mathbf{C}_{\text{e}}}:\mathbf{X} =
  \mathbf{X} - \frac{1}{3}\text{tr}(\mathbf{C}_{\text{e}}\mathbf{X})\mathbf{C}^{-1}_{\text{e}}, \quad
  \text{for all} \ \mathbf{X} \in Sym.
 \end{equation}
Finally, the second Piola-Kirchhoff operating on the reference configuration is computed through
 \begin{equation}
  \tilde{\mathbf{T}} = \mathbf{F}^{-1}_{\text{i}} \hat{\mathbf{S}} \mathbf{F}^{-\text{T}}_{\text{i}}
  = \frac{2 \rho_{R} f}{ \lambda^{2}_{\text{i}}} \mathbb{P}_{\mathbf{C}}:\mathbf{M}.
 \end{equation}
The Clausius-Duhem inequality is reduced to
\begin{equation}
    \delta_{\text{i}} = \frac{1}{2\rho_{R}} \tilde{\mathbf{T}}:\dot{\mathbf{C}} - \dot{\Psi}_{fiber}^{visc} \ge 0.
\end{equation}
Further, using the Mandel tensor $\hat{\mathbf{C}}_{\text{e}}\hat{\mathbf{S}}$, we
obtain the well-known expression for the mechanical dissipation:
\begin{equation}
    \delta_{\text{i}} = \frac{1}{\rho_{R}} (\hat{\mathbf{C}}_{\text{e}}\hat{\mathbf{S}}):\hat{\mathbf{L}}_{\text{i}} \ge 0.
\end{equation}
Using \eqref{fiberElas} and \eqref{fiberElas2} we see that the Mandel tensor is non-symmetric:
\begin{equation}
    \hat{\mathbf{C}}_{\text{e}}\hat{\mathbf{S}} = 2\rho_{R}f(\hat{\mathbf{C}}_{\text{e}}\mathbf{M})^{\text{D}} \notin Sym,
\end{equation}
which is typical for anisotropic strain energy functions.

For what follows it is useful to check that
\begin{equation}
    \displaystyle \text{tr}(\mathbf{M}\mathbf{C}_{\text{e}}\mathbf{L}_{\text{i}}) = \frac{ \dot{\lambda}_{\text{i}} \lambda^{2} }{ \lambda^{3}_{\text{i}} }.
\end{equation}
Employing this relation we obtain the mechanical dissipation in the form:
 \begin{equation}
    \delta_{\text{i}} = 2 f \frac{ \displaystyle \dot{\lambda}_{\text{i}} \lambda^{2} }{ \displaystyle \lambda^{3}_{\text{i}} } \ge 0.
 \end{equation}
Assuming that the material is elastically stable ($f >0$), the Clausius-Duhem inequality reduces to
\begin{equation}\label{Duhem2}
    f \cdot \dot{\lambda}_{\text{i}} \ge 0.
\end{equation}
It requires that the fibers elongate under tension and contract under compression.

Now we postulate the flow rule satisfying \eqref{Duhem2} considering that under uniaxial loading the model
must reproduce the following type of uniaxial material behaviour in terms of the inelastic
logarithmic strain $\varepsilon^{(log)}_{\text{i}} := \ln(\lambda_{\text{i}})$ and the true stress $\sigma$:
\begin{equation}
\dot{\varepsilon}^{(log)}_{\text{i}}     = \frac{1}{\displaystyle 2\eta} \sigma.
\end{equation}
Thus, we obtain
\begin{equation}\label{EvolutionFiberInel}
  \frac{\displaystyle \dot{\lambda_{\text{i}}}}{ \displaystyle \lambda_{\text{i}} } = \frac{1}{\displaystyle\eta}
  f\Big( \big(\frac{\displaystyle \lambda}{\displaystyle \lambda_{\text{i}}}\big)^{2}\Big)
    \cdot \frac{\displaystyle \lambda^{2}}{\displaystyle \lambda^{2}_{\text{i}}} \rho_{R}.
\end{equation}
This flow rule can be formulated in terms of the elastic stretch as well
\begin{equation} \label{EvolutionFiber}
\begin{array}{l}
    \frac{\displaystyle \dot{\lambda_{\text{e}}}}{\displaystyle \lambda_{\text{e}}} =
     -\frac{\displaystyle \rho_{R}}{\displaystyle \eta} f( \lambda_{\text{e}}^{2}) \lambda_{\text{e}}^{2}.
\end{array}
\end{equation}
Note that both formulations of the flow rule are one dimensional.

Any type of the fiber potential discussed in Section 2.2 can be implemented
by substituting $\lambda_{\text{e}}$ in place of $\lambda$.
To be definite, as $\Psi_{fiber}^{visc}$ we take the Holzapfet potential \eqref{HolzapfelAniso}:
\begin{equation} \label{PsiHolzapfel}
\begin{array}{l}
    \Psi_{Holzapfel}(\lambda_{\text{e}}^{2}) = \frac{\displaystyle
    k_{1}}{\displaystyle  k_{2}} ( e^{ k_{2}(\lambda_{\text{e}}^{2} - 1)^{2} } - 1), \quad
    f = 2k_{1}(\lambda_{\text{e}}^{2} - 1)e^{ k_{2}(\lambda_{\text{e}}^{2} - 1)^{2} }.
\end{array}
\end{equation}

Note that, in contrast to the previous subsection, the elastic properties are anisotropic here.
For such a type of stress response, the inelastic spin need to be properly defined. Indeed, for the transversely
isotropic material considered here, the spin is restricted by the
assumption $ \mathbf{F}_{\text{i}} \tilde{\mathbf{a}} = \lambda_{\text{i}} \tilde{\mathbf{a}}$.

\section{Time stepping methods}

The evolution equations $\eqref{Cinelast}_4$ and \eqref{EvolutionFiberInel} are stiff. Therefore,
explicit time stepping would yield poor results. In this section we discuss their implicit integration.

\subsection{Explicit update formula for isotropic Maxwell body}
In order to solve the evolution equation $\eqref{Cinelast}_3$ pertaining to the Maxwell body in case of the neo-Hookean potential
 \eqref{IsotrNeoHook} we can use an efficient iteration-free algorithm, which was proposed by Shutov et al. in \cite{Shutov2013}.
Consider a generic time step from $t_{n}$ to $t_{n+1}$; assume that the current deformation
gradient ${}^{n+1}\mathbf{F}$ and the previous inelastic Cauchy-Green tensor ${}^{n}\mathbf{C}_{\text{i}}$ are known.
The Euler backward method for  $\eqref{Cinelast}_3$ takes the form:
 \begin{equation} \label{CinEuler}
        {}^{n+1}\mathbf{C}^{\text{EBM}}_{\text{i}} = {}^{n}\mathbf{C}_{\text{i}} + \frac{\triangle t \mu}{\eta} ({}^{n+1}\overline{\mathbf{C}} ({}^{n+1}\mathbf{C}^{\text{EBM}}_{\text{i}} )^{-1} )^{D} \cdot {}^{n+1}\mathbf{C}^{\text{EBM}}_{\text{i}},
\end{equation}
where $ {}^{n+1}\mathbf{C}^{\text{EBM}}_{\text{i}} $ stands for the EBM-solution at $t_{n+1}$. Abbreviating
 \begin{equation}
 \beta := \frac{1}{3} \frac{\triangle t \mu}{\eta} \text{tr}({}^{n+1}\overline{\mathbf{C}} ({}^{n+1}\mathbf{C}^{\text{EBM}}_{\text{i}} )^{-1} ),
 \end{equation}
 we re-write \eqref{CinEuler} as follows
\begin{equation} \label{CinEuler2}
 {}^{n+1}\mathbf{C}^{\text{EBM}}_{\text{i}} = {}^{n}\mathbf{C}_{\text{i}} + \frac{\triangle t \mu}{\eta} {}^{n+1}\overline{\mathbf{C}} - \beta {}^{n+1}\mathbf{C}^{\text{EBM}}_{\text{i}}.
\end{equation}
This yields
 \begin{equation} \label{CinEuler3}
        {}^{n+1}\mathbf{C}^{\text{EBM}}_{\text{i}} = \frac{1}{1+\beta}({}^{n}\mathbf{C}_{\text{i}} + \frac{\triangle t \mu}{\eta} {}^{n+1}\overline{\mathbf{C}}).
\end{equation}
The scalar $\beta$ is unknown, since it depends on the unknown solution. On the other hand,
for any second-rank tensor $\mathbf{A}$ such that $\det \mathbf{A} >0$, we have
 \begin{equation} \label{Aequation}
\overline{ \frac{1}{1+\beta} \mathbf{A}} = \overline{\mathbf{A}}.
\end{equation}
Thus, applying the operation $\overline{(\cdot)}$ to the both sides of \eqref{CinEuler3}, we have
\begin{equation}\label{CinEuler3}
{}^{n+1}\mathbf{C}_{\text{i}} = \overline{{}^{n+1}\mathbf{C}^{\text{EBM}}_{\text{i}}} = \overline{{}^{n}\mathbf{C}_{\text{i}} + \frac{\triangle t \mu}{\eta} {}^{n+1}\overline{\mathbf{C}}}.
\end{equation}
This is an explicit update formula reported in \cite{Shutov2013}. Since it is iteration free, it is especially robust. The algorithm is
first order accurate and unconditionally stable \cite{Shutov2013}.
Concerning the accuracy, the algorithm is equivalent to the Euler backward method with a subsequent correction of the incompressibility and to the
implicit methods based on the tensor exponent \cite{Shutov2013}.
Thanks to the exact incompressibility, the algorithm allows one to suppress the accumulation
of the numerical error \cite{Shutov2010}.
As shown in \cite{Shutov2013}, the method exactly preserves the w-invariance of the constitutive equations under
isochoric change of the reference configuration (for a general definition of the w-invariance
the reader is referred to \cite{Shutov2014}).

Note that this explicit update formula exploits the special structure
of the neo-Hookean potential \eqref{IsotrNeoHook}.
An iteration-free time stepping algorithm dealing
with a more general storage energy function of the Mooney-Rivlin type
was recently proposed in \cite{Shutov2017}.
Another simple generalization to the case of the Yeoh potential was discussed in \cite{LandgrafShutov}.

\subsection{Efficient algorithm for the fiber-like Maxwell body}

\begin{figure}\centering
    \psfrag{A}[m][][1][0]{$ {}^{n}\lambda_{e}$}
    \psfrag{B}[m][][1][0]{$ {}^{n+1}\lambda_{e}$}
    \scalebox{1.0}{\includegraphics{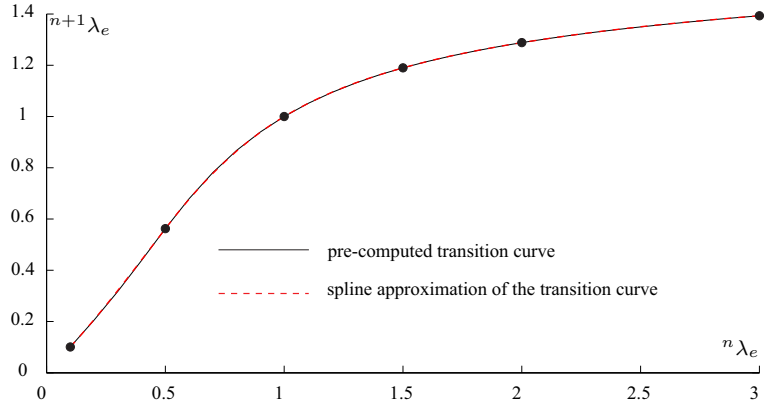}}
    \caption{Transition curve providing ${}^{n+1} \lambda_{\text{e}}$ as a function of ${}^{n} \lambda_{\text{e}}$
    within a single time step.
    The following   parameters were used for this particular curve: $\Delta t = 2^{-7}$ s, $\eta = 5$ KPa $\cdot    $ s,
  $k_1 = 130$ KPa, $k_2 = 0.5$. \label{SketchSpline}}
\end{figure}

Now let us consider the Euler backward method for the evolution equation $\eqref{EvolutionFiber}$
governing the fiber viscosity:
\begin{equation} \label{EulerFiberLambdaE}
    {{}^{n+1}\lambda_{e}} = {{}^{n}\lambda_{e}} - \frac{\triangle t \rho_{R} }
    {\eta} f( {}^{n+1}\lambda_{e}^{2} ) \ {}^{n+1}\lambda_{e}^{3}.
\end{equation}
Solving it by the classical Newton method we obtain a function
 ${{}^{n+1}\lambda_{e}} = {{}^{n+1}\lambda_{e}} ({{}^{n}\lambda_{e}})$.
In the following we assume that  ${\lambda_{e}}$ remains in the interval $[0.1,3]$.
In order to improve the efficiency of the numerical algorithm we subdivide the
  interval into five subintervals: $ [0.1,0.5], [0.5,1], [1,1.5], [1.5,2], [2,3]$
  and interpolate the ${{}^{n+1}\lambda_{e}} ({{}^{n}\lambda_{e}})$ function by the cubic spline
  $\text{Spline} ({{}^{n}\lambda_{e}})$
  using the mentioned key points. An example of such an interpolation is shown in Figure \ref{SketchSpline},
  which corresponds to the Holzapfel potential $\eqref{PsiHolzapfel}$.

Using the pre-computed spline we can solve (\ref{EulerFiberLambdaE}) approximately, by evaluating the function
 $ {{}^{n+1}\lambda^{\ast}_{e}} = \text{Spline} ({{}^{n}\lambda_{e}}) $. Afterwards, in order to
 improve the accuracy we perform a single Newton iteration finally yielding  ${{}^{n+1}\lambda_{e}}$.
 Thus, this method requires only one Newton iteration and in can be used for any regular function $f$.

\subsection{Testing the algorithm for fiber viscoelasticity}

\begin{figure}\centering
    \psfrag{A}[m][][1][0]{$t, s$}
    \psfrag{B}[m][][1][0]{$\varepsilon = \ln\frac{l}{l_{0}}$}
    \scalebox{1.0}{
    \includegraphics{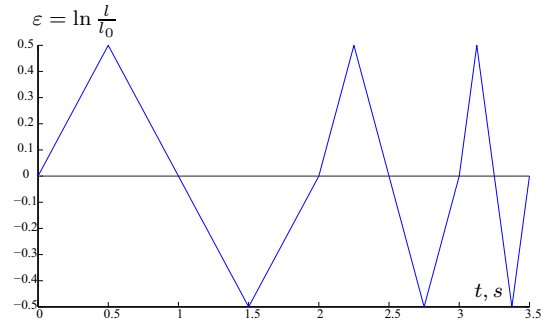}}
    \caption{Prescribed dependence of the logarithmic strain $\varepsilon$ of the sample on time $t$. \label{lambdaGraph}}
\end{figure}

\begin{figure}\centering
    \psfrag{A}[m][][1][0]{$t, s$}
    \psfrag{B}[m][][1][0]{$\sigma$, KPa}
    \scalebox{1.0}{
    \includegraphics{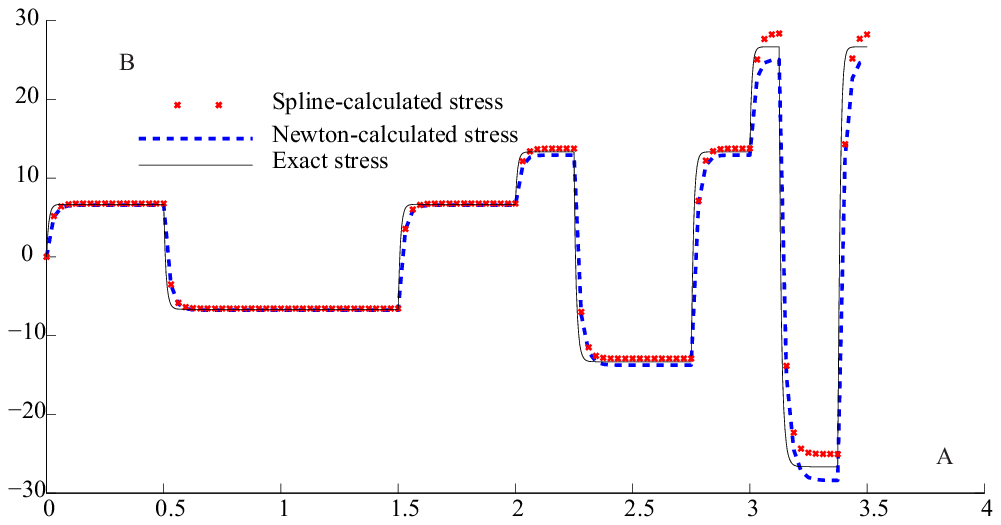}}
    \caption{Computed stress history for the time step size $\triangle t = 2^{-5} s.$ \label{Stress5}}
\end{figure}

\begin{figure}\centering
    \psfrag{A}[m][][1][0]{$t, s$}
    \psfrag{B}[m][][1][0]{$\sigma$, KPa}
    \scalebox{1.0}{
    \includegraphics{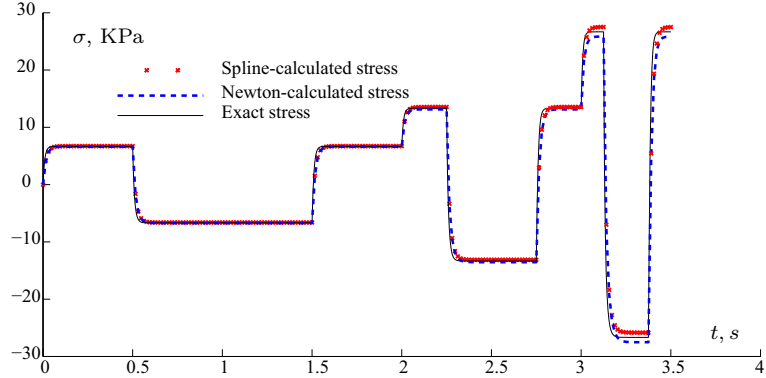}}
    \caption{Computed stress history for the time step size $\triangle t = 2^{-6} s.$ \label{Stress6}}
\end{figure}

In order to test the convergence of the newly proposed algorithm,
a series of numerical experiments is carried out using the Holzapfel potential \eqref{PsiHolzapfel}.
In these tests a single fiber is subjected to a uniaxial isochoric strain-controlled non-monotonic loading.
We set $\mathbf{F} = \lambda \tilde{\mathbf{a}} \otimes \tilde{\mathbf{a}} +
\lambda^{-1/2} (\mathbf{1} - \tilde{\mathbf{a}} \otimes \tilde{\mathbf{a}}) $, where
$\tilde{\mathbf{a}} $ is the unit vector defining the fiber orientation in the reference configuration and
$\lambda = l/l_0$ is the prescribed stretch. The loading program is show
in Figure \ref{lambdaGraph}.
The employed material parameters are summarized in Table \ref{ParamHolzapfel}.
The numerical solution obtained using the classical Euler backward in combination with the Newton iteration and extremely
small time step size $\triangle t = 10^{-6} s$ is considered to be exact and denoted by $\sigma^{exact}$.
The quantity of interest is the axial stress $\sigma = \mathbf{T} : (\tilde{\mathbf{a}} \otimes \tilde{\mathbf{a}}) $, where
$\mathbf{T}$ stands for the true stress.
For step sizes $\triangle t \in \{2^{-5} s, 2^{-6} s \}$ we present the corresponding
stress histories obtained by the classical EBM and the novel method, see Figures \ref{Stress5} and \ref{Stress6}.
Both methods accurately describe the stress response under the smooth loading;
the error increases shortly after the strain rate experiences a jump.

\begin{table}[H]
\caption{Set of material parameters of a single fiber used for accuracy testing}
\label{ParamHolzapfel}
\begin{center}
\begin{tabular}{c c c}
\hline
$k_{1}$ & $k_{2}$ & $\eta$ \\
\hline
130 KPa & 0.5 &   5 KPa $\cdot s $ \\
\hline
\end{tabular}
\end{center}
\end{table}

\begin{figure}\centering
    \psfrag{A}[m][][1][0]{$t, s$}
    \psfrag{B}[m][][1][0]{$\triangle\sigma$, KPa}
    \scalebox{1.0}{
    \includegraphics{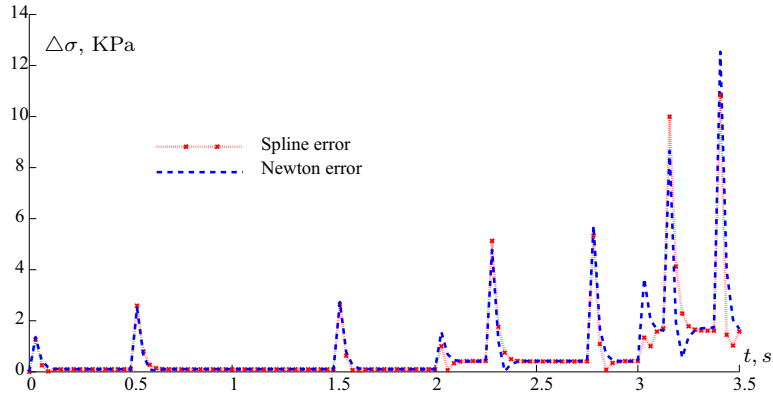}}
    \caption{Numerical error for the time step size $\triangle t = 2^{-5} s$. \label{Error5}}
\end{figure}

\begin{figure}\centering
    \psfrag{A}[m][][1][0]{$t, s$}
    \psfrag{B}[m][][1][0]{$\triangle\sigma$, KPa}
    \scalebox{1.0}{
    \includegraphics{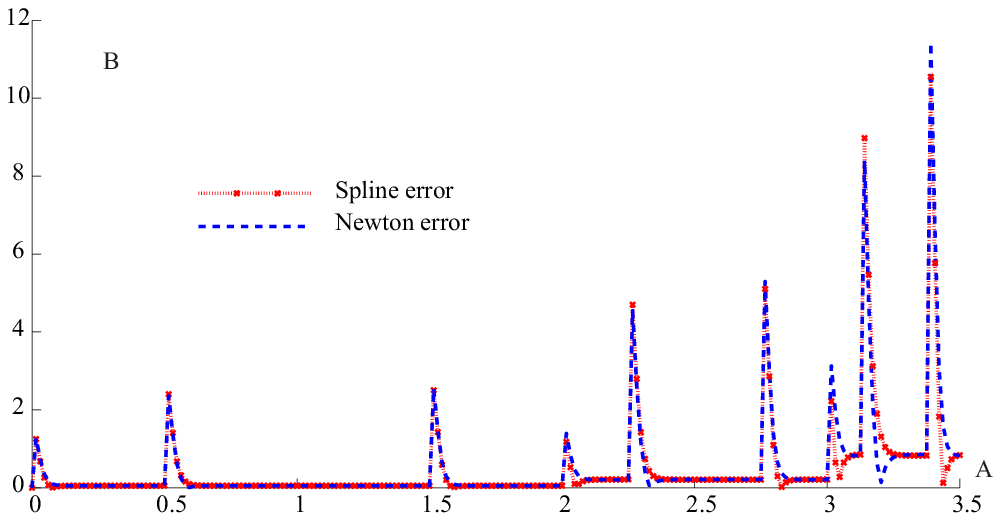}}
    \caption{Numerical error for the time step size  $\triangle t = 2^{-6} s.$ \label{Error6}}
\end{figure}

\begin{figure}\centering
    \psfrag{A}[m][][1][0]{$t, s$}
    \psfrag{B}[m][][1][0]{$\triangle\sigma$, KPa}
    \scalebox{1.0}{
    \includegraphics{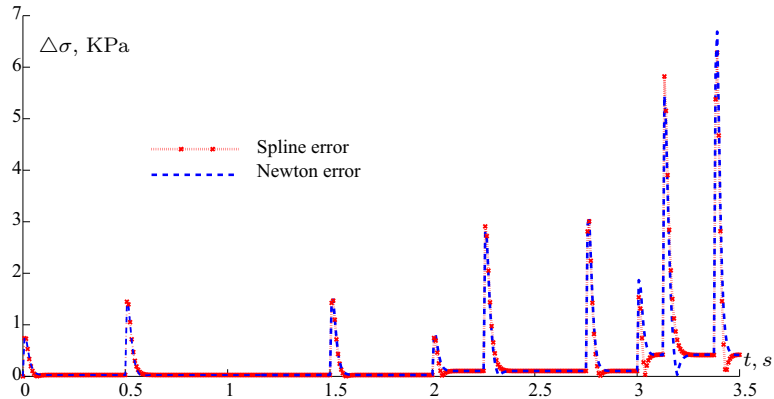}}
    \caption{Numerical error for the time step size  $\triangle t = 2^{-7} s.$ \label{Error7}}
\end{figure}

The deviation of the numerical solution from the exact solution is denoted by $\Delta \sigma :=  |\sigma^{exact} - \sigma^{numerical}|  $.
As can be seen from Figures \ref{Error5}, \ref{Error6}, \ref{Error7},
the classical method with the Newton iterations and the novel method with pre-computed splines exhibit the same accuracy
for the time step sizes $\triangle t \in \{2^{-5} s, 2^{-6} s, 2^{-7} s \}$.

\section{Demonstration problem: pressurized composite tube}

\begin{figure}\centering
    \psfrag{in}[m][][1][0]{$7.5$}
    \psfrag{mid}[m][][1][0]{$8.0$}
    \psfrag{out}[m][][1][0]{$2.5$}
    \psfrag{Ri}[m][][1][0]{$R_{\text{i}} = 100.0$}
    \psfrag{Ro}[m][][1][0]{$R_{o} = 118.0$}
    \psfrag{g1}[m][][1][0]{$2 \gamma$}
    \psfrag{g2}[m][][1][0]{$2 \gamma$}
    \scalebox{0.9}{
    \includegraphics[width=0.6\textwidth]{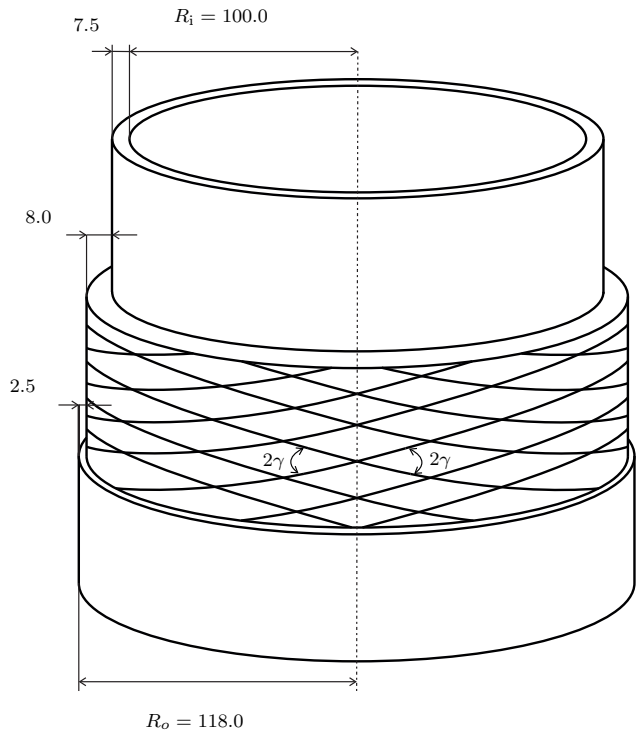}}
    \caption{Dimensions of the composite tube and its structure.
    The internal and external layers are isotropic; the
    middle layer corresponds to a material reinforced by
    two families of fibers (cf. \cite{Holzapfel2001}). \label{StrInvComposite}}
\end{figure}

In order to demonstrate the efficiency of the advocated multiplicative approach to the modelling of fiber-reinforced
viscoelastic composites, we solve here an initial boundary value problem.
This problem was analyzed previously by Holzapfel et al. in \cite{Holzapfel2001}.
Let us consider a composite tube of circular cross section, comprising three layers: The internal and external layers are isotropic;
 the middle layer is made of a composite, reinforced by two families of fibers, as shown in Figure \ref{StrInvComposite}.
 By $\mathbf{a}_{1}$ and $\mathbf{a}_{2}$ denote the unit fiber orientation vectors pertaining to these two families.
 The tube is loaded by internal pressure $P$. To account for the pressure exerted on the plugs at both ends of the tube,
 an axial force $F = P \times \pi \times (\text{inner radius})^2$ is applied.
The material is assumed to be incompressible. This allows us to solve the initial boundary value problem
without using FEM by a semi-analytical procedure (details are presented in Appendix A).

\subsection{Purely hyperelastic case \label{HyperElast}}

Viscous effects are neglected in this subsection. Following the iso-strain approach (cf. \cite{Holzapfel2001}), we
assume that the Helmholz free energy of the composite material
is decomposed into the following sum:
\begin{equation}\label{PsiFunctionComposite}
    \Psi = \Psi_{MR}(\overline{\mathbf{C}}) + \Psi_{fiber}((\lambda^{\mathbf{a}_{1}})^{2})
    + \Psi_{fiber}((\lambda^{\mathbf{a}_{2}})^{2}),
\end{equation}
where
\begin{equation}\label{PsiFunctionComposite2}
(\lambda^{\mathbf{a}_{k}})^{2} = \overline{\mathbf{C}}:( \mathbf{a}_{k} \otimes \mathbf{a}_{k}) = \mathbf{a}_{k} \cdot \overline{\mathbf{C}} \cdot \mathbf{a}_{k},
\quad \text{for} \ k \in \{1,2\}.
\end{equation}
The material parameters for each layer are taken from \cite{Holzapfel2001} and summarized in Table \ref{ParamHolzapfel}.

\begin{table}[H]
\caption{Set of material parameters in the purely hyperelastic case taken from \cite{Holzapfel2001}}
\label{ParamCompositeTube}
\begin{center}
\begin{tabular}{ l c c c }
\hline
  & Inner layer & Middle layer & External layer\\
  \hline
$c_{1}$ [KPa]      & 4.0  & 0.86   & 4.0 \\
$c_{2}$ [KPa]      & 1.0  & 0.215  & 1.0 \\
$k_{1}$ [KPa]      & 0.0  & 260.0  & 0.0 \\
$k_{2}$ [KPa]      & 0.0  & 0.5    & 0.0 \\
$\gamma$ [$^{\circ}$] & -    & 33.1   & -\\
\hline
\end{tabular}
\end{center}
\end{table}

Simulated and experimental values of the axial stretch $\lambda_{z}$ and the hoop stretch $\lambda_{\theta}$
are plotted versus the applied inner pressure $P$ in Figure \ref{StrInvCheck};
experimental data are taken from the work of Wiesemann \cite{Wiesemann1995}.
Our simulation results coincide with the results obtained previously
by Holzapfel et al. in \cite{Holzapfel2001} using the FEM.
As mentioned before, in the current study the boundary value problem is solved without resorting to FEM by
the semi-analytical method.
Moreover, the presented material is implemented in MSC.MARC and a very good correspondence between
the semi-analytical solution and the FEM solution is obtained, see Appendix B.

Note that for small pressure the tube\emph{ tends to reduce its diameter as the pressure increases} (see Figure \ref{StrInvCheck}).
This counterintuitive response is known as the \emph{stretch inversion phenomenon}.
As is seen from Figure \ref{StrInvCheck}, the stretch inversion phenomenon is captured by the considered idealization
of the fiber-reinforced structure. Within the current model, its explanation is as follows.
At the initial stage of pressure growth, the internal volume is most efficiently increased by a rapid elongation of the tube which
is accompanied by alignment of reinforcing fibers along the axis. During that stage, the tube's diameter decreases.
After a certain rotation of the fibers, this mechanism is not efficient anymore and the volume growth is realized
by a nearly uniform elongation in axial and hoop directions. Thus, the stretch inversion phenomenon is essentially geometrically nonlinear
and it can not be described by theories with linear kinematics.

\subsection{Identification of parameters $\gamma$ and $k_{1}$ \label{FirstSearch}}

\begin{figure}\centering
    \psfrag{P}[m][][1][0]{$P$, KPa}
    \psfrag{P2}[m][][1][0]{$P$, KPa}
    \psfrag{Lamt}[m][][1][0]{$\lambda_{\theta}$}
    \psfrag{Lamz}[m][][1][0]{$\lambda_{z}$}
    \scalebox{1.0}{
    \includegraphics{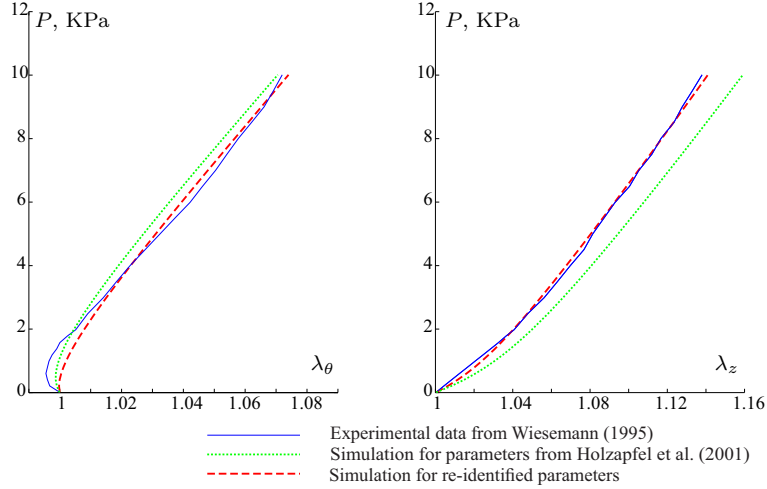}}
    \caption{Inflation of a pressurized composite tube: experimental data
    from \cite{Wiesemann1995} and simulation results using different sets of parameters. \label{StrInvCheck}}
\end{figure}

According to Table \ref{ParamHolzapfel},
the main load is carried by the fiber-reinforced middle layer.
Therefore, in the following we focus on a more accurate identification of its parameters
 basing on the available experimental data.
The problem is as follows: having the experimental curves $\lambda_{z}^{\text{exp}}(P)$ and $\lambda_{\theta}^{\text{exp}}(P)$ find
 $\gamma$ and $k_{1}$, providing the simulated response most close to the experimental one. Here, the parameter $k_{2}$ is fixed, since
 it becomes important at large stretches only and the available data correspond to stretches below 10 \%.
 Thus, we set $k_{2}=0.5$. For the parameter identification we build the following error functional:

\begin{equation}
    \Phi(P_{\text{i}}) = \sum_{i=1}^{N} ( \lambda_{\theta}^{exp}(P_{\text{i}}) - \lambda_{\theta}^{num}(P_{\text{i}}))^{2} +
    \sum_{i=1}^{N} ( \lambda_{z}^{exp}(P_{\text{i}}) - \lambda_{z}^{num}(P_{\text{i}}))^{2},
\end{equation}
where $N$ is the number of experimental points; $ \lambda^{exp}(P_{\text{i}})$ and $\lambda^{num}(P_{\text{i}})$ are the experimental and numerical stretches
corresponding to the $i$-th point. The error functional is minimized using the Levenberg-Marquardt method;
the identified values are as follows: $k_{1}^{*} = 267.5 \ \text{KPa}, \gamma^{*} = 0.5871 \ \text{rad} \approx 33,638^{\circ}$. Simulation
results obtained for the identified parameters are shown in Figure \ref{StrInvCheck}. Although the identified parameters $\gamma$ and $k_{1}$
are close to the parameters reported in \cite{Holzapfel2001} (cf. Table \ref{ParamHolzapfel}), the newly simulated curves are closer to the experimental data.

The stress distribution through the thickness of the tube is calculated
according to the procedure summarized in Appendix A; the results are shown in Figure \ref{StressDistr}.
Note that the hoop stresses are negative within the internal layer.
This effect is caused by the fact that the internal layer is restrained by a more rigid middle layer. This
leads to occurrence of essential hydrostatic compression within the inner layer.
The occurrence of negative hoop stresses within the inner layer induces additional load on the middle layer.

 \begin{figure}\centering
    \psfrag{Sigma}[m][][1][0]{Stress, KPa}
    \psfrag{dR}[m][][1][0]{$r-r_{\text{i}}$, mm}
    \psfrag{SigmaR}[m][][1][0]{$\sigma_{rr}$}
    \psfrag{SigmaT}[m][][1][0]{$\sigma_{\theta\theta}$}
    \psfrag{SigmaZ}[m][][1][0]{$\sigma_{zz}$}
    \scalebox{1.0}{
    \includegraphics{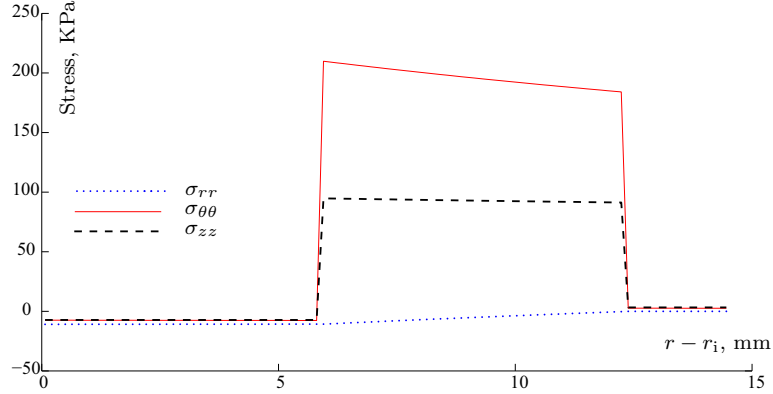}}
    \caption{Distribution of true stresses in the radial, hoop, and axial directions
    obtained for the inner pressure $P$=10 KPa \label{StressDistr}.}
\end{figure}

\subsection{Viscoelastic case}

Viscous effects are crucial in numerous engineering and medical applications. In order to demonstrate
the applicability of the advocated approach we introduced viscous effects. Following
the iso-strain approach, we postulate
\begin{equation}\label{PsiFunctionComposite}
    \Psi = \Psi_{MR}(\overline{\mathbf{C}}) + \Psi_{fiber}((\lambda^{(\mathbf{a}_{1})})^{2}) + \Psi_{fiber}((\lambda^{(\mathbf{a}_{2})})^{2}) +
    \Psi_{neo-Hooke}(\overline{\mathbf{C}}_{e}) + \Psi_{fiber}^{visc}((\lambda_{e}^{(\mathbf{a}_{1})})^{2}) + \Psi_{fiber}^{visc}((\lambda_{e}^{(\mathbf{a}_{2})})^{2}),
\end{equation}
where $\overline{\mathbf{C}}_{e}$ is introduced in Subsection \ref{Ce}; $\lambda_{e}^{(\mathbf{a}_{\text{i}})}$ is defined in Subsection 2.4 (cf. \eqref{lambdaei});
for the fiber-related Maxwell body we assume $k_{2}^{vis} = k_{2}$.

\begin{figure}\centering
    \psfrag{P}[m][][1][0]{$P$, KPa}
    \psfrag{P2}[m][][1][0]{$P$, KPa}
    \psfrag{Lamt}[m][][1][0]{$\lambda_{\theta}$}
    \psfrag{Lamz}[m][][1][0]{$\lambda_{z}$}
    \psfrag{Reid}[m][][1][0]{Расчет для полученных параметров}
    \psfrag{Mod}[m][][1][0]{Синтетические данные с внесенной ошибкой}
    \scalebox{1.0}{
    \includegraphics{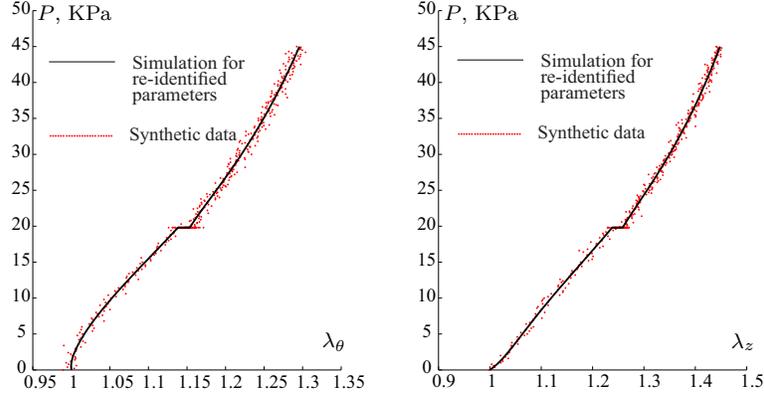}}
    \caption{Inflation of a pressurized visco-elastic composite tube: stretch in the hoop direction (left)
    and axial direction (right) as a function of the applied pressure. Synthetic data and
    simulation results using re-identified parameters are shown. \label{StrInvViscGraph}}
\end{figure}

As in the previous subsection we assume that the material parameters governing the isotropic part of the stress response are known
everywhere. Thus, its remains to identify the parameters of the fibers in the middle layer: orientation angle $\gamma$, stiffness $k_{1}$,
nonlinearity $k_{2}$,
viscosity of the Maxwell body $\eta_{fib}$, stiffness $k_{1}^{vis}$ of the Maxwell body.
In order to demonstrate that these unknown parameters can be identified using real experimental data on the loading of the tube,
we consider the following re-identification
problem. First, we manually preset certain values of the material parameters, see Table \ref{ParamsTable}.
Next, we carry out a simulation where the applied pressure linearly grows from 0 KPa to 20 KPa within 0.1 seconds.
Then the pressure is held fixed for 0.1 seconds. Finally, the pressure grows from 20 KPa to 45 KPa within 0.25 seconds.
In order to mimic the real experimental data, the simulation results are spoiled by a stochastic noise, thus
producing synthetic data:

\begin{equation}
    {\lambda^{\text{synth}}_{\theta \text{i}}} = \lambda_{\theta \text{i}}^{\text{simulated}} + Noise^{\theta}_{\text{i}}, \quad
    {\lambda^{\text{synth}}_{z \text{i}}}= \lambda_{z \text{i}}^{\text{simulated}} + Noise^{z}_{\text{i}},
\end{equation}
where $Noise^{x}_{\text{i}}$ is a random variable with normal distribution of zero mean
 and  $5 \cdot 10^{-3} \max(\lambda_{x})$ variance, $x \in \{\theta, z\}$.
These synthetic data are shown by dots in Figure \ref{StrInvViscGraph}. Next, in order to re-identify the material
parameters, a least square error functional is minimized using the Levenberg-Marquardt method.
Figure \ref{StrInvViscGraph} shows the comparison of the simulation results using re-identified parameters and the synthetic data.
The original and re-identified parameters are summarized in Table \ref{ParamsTable}.
Although the synthetic data are highly noisy, \emph{five material parameters
can be accurately identified using the single test}.

\begin{table}[H]
\caption{Original and re-identified parameters governing the middle layer of the composite tube.}
\label{ParamsTable}
\begin{center}
\begin{tabular}{ c c c }
 \hline
  & Original & Re-identified \\
  \hline
$\gamma$ & 33.10  & 33.10\\
$k_{1}$ & 260.0 KPa & 261.2 KPa\\
$k_{2}$ & 0.5 & 0.4898\\
$k_{1}^{vis}$ & 130.0 KPa & 133.5 KPa\\
$\eta_{fib}$ & 10.0 & 9.880\\
\hline
\end{tabular}
\end{center}
\end{table}

\section{Discussion and conclusion}

A geometrically exact approach to the modelling of fiber-reinforced viscoelastic
structures is considered. The overall model of the composite material is obtained using the
iso-strain assumption, thus
yielding a low cost computational
tool suitable for large scale analysis. In the current paper, formulations of the Maxwell body based on the multiplicative
decomposition of the deformation gradient (Sidoroff decomposition) are advocated.
Using the multiplicaitve split, it is possible to create models which are objective and thermodynamically consistent.
Moreover, very simple iteration-free algorithms are available for the efficient implementation of these models.
In the current study each viscous element is modelled using a single Maxwell branch.
A complex relaxation spectrum in a broad frequency range can be accurately
described using a number of such brunches.

In some cases, the existence of pre-stresses needs to be accounted for \cite{Vaishnav1983}. The pre-stresses
can be easily introduced within the multiplicative approach. Even more,
for a flexible modelling of pre-stressed components, different constituents
of the model may exhibit different unstressed configurations (cf. the so-called constrained mixture theory
\cite{Humphrey2002}, \cite{Cyron2016}).
An important property of the Maxwell formulations based on the multiplicative decomposition (Sections 2.3 and 2.4)
is that the corresponding evolution equations are independent of the choice of the
reference configuration \cite{Shutov2014}, \cite{Shutov2016}. This invariance
property simplifies the numerical modelling essentially.

For a better description of the stress response under local compression, a new family of
energy storage functions (hyperelastic potentials) is suggested.
Using the additional parameter (cf. equation \eqref{GeneralizedPotential}),
one may account for the interaction between fiber and matrix under compression.
Under tensile stresses, these potentials reproduce the well-known potential of Holzapfel et al.

As a demonstration problem, inflation of a pressurized composite tube is analyzed.
Due to the assumed incompressibility of the material, the initial boundary value problem can be solved
without resorting to FEM by a semi-analytical procedure.
The obtained solutions are valuable since they can serve as a
benchmark for FEM implementations of the composite model.
An interesting result is that five parameters of the composite model
can be identified with a good accuracy using a single test on the inflation of the tube (cf. Table \ref{ParamsTable}).
The numerical tests did not reveal any unphysical effects. Thus, the framework is applicable to
the analysis of large deformations of fiber-reinforced composite structures.

The results regarding efficient numerics, accurate description of the
global structural response, and reliable parameter identification show the suitability
of the chosen approach to fiber-reinforced composites.

\noindent \textbf{Acknowledgments} The financial support provided by the RFBR (grant number 17-08-01020) and by
the integration project of SB RAS is acknowledged.

\noindent \textbf{Compliance with ethical standards}

\noindent \textbf{Conflict of interest} The authors declare that they have no
conflict of interest.

\section*{Appendix A: Semi-analytical procedure for the initial boundary value problem}

Here we describe a semi-analytical procedure used to solve the initial boundary value problem.
First, we introduce a cylindrical coordinate system $(r,\theta, z)$ in the current configuration and
$(R, \Theta, Z)$ in the reference configuration.
Let $\lambda_{r}$, $\lambda_{\theta}$, and $\lambda_{z}$ be the stretches
in the radial, hoop, and axial directions, respectively.
The local incompressibility condition reads
 \begin{equation}
 \lambda_{r}\cdot\lambda_{\theta}\cdot\lambda_{z} = 1.
\end{equation}
Assume that $L$, $R_{\text{o}}$, and $R_{\text{i}}$ are the tube's original length, outer radius
and inner radius, respectively; $l$, $r_{\text{o}}$, and $r_{\text{i}}$
are the corresponding dimensions in the current configuration.
During the deformation, the domain $R \in [R_{\text{i}}, \hat{R}] $, $Z \in [0, \hat{Z}]$
transforms to the domain $r \in [r_{\text{i}}, \hat{r}]$, $z \in [0, \hat{z}]$.
Due to the incompressibility, its volume remains constant and thus we have
\begin{equation}
    \displaystyle \hat{r} = \sqrt{\frac{\hat{R}^{2}-R^{2}_{\text{i}}}{\lambda_{z}}+r^{2}_{\text{i}}}, \quad \hat{z} = \lambda_{z} \hat{Z},
\end{equation}
\begin{equation}
    \displaystyle \lambda_{\theta} = \frac{\hat{r}}{\hat{R}}\frac{\partial \hat{\theta}}{\partial \hat{\Theta}} =
    \frac{\hat{r}}{\hat{R}}, \quad \lambda_{r} = \frac{\hat{R}}{\hat{r}\lambda_{z}}.
\end{equation}
By $\lambda_{\theta}^{\text{inner}}$ denote $\lambda_{\theta}$ on the inner surface
of the tube. The incompressibility condition allows us to describe the entire kinematics by only two scalar
quantities: $\lambda_{z}$ and $\lambda_{\theta}^{\text{inner}}$.

Taking the cylindrical symmetry of the problem into account, the deformation gradient takes the form:
\begin{equation}\label{DefGradF}
    \displaystyle \mathbf{F} = \lambda_{r} \mathbf{e}_r \otimes \mathbf{e}_r + \lambda_{\theta} \mathbf{e}_{\theta} \otimes \mathbf{e}_{\theta}
+  \lambda_{z} \mathbf{e}_{z} \otimes \mathbf{e}_{z}.
\end{equation}
Since the tube geometry, material properties, and applied loads are independent of the axial and hoop coordinates,
only a single equilibrium equation needs to be satisfied:
\begin{equation}\label{requibl}
    \displaystyle \frac{d\mathbf{T}_{rr}}{dr} + \frac{\mathbf{T}_{rr} - \mathbf{T}_{\theta \theta}}{r} = 0.
\end{equation}
Assuming that the outer surface is stress free, we have
\begin{equation}
     \mathbf{T}_{rr}\mid_{r = r_{\text{o}}} = 0 \Longrightarrow
     \mathbf{T}_{rr}(\xi) = \int_{\xi}^{r_{\text{o}}} \frac{\mathbf{T}_{rr} - \mathbf{T}_{\theta \theta}}{r} dr.
\end{equation}
Thus we obtain the following expression for the pressure on the inner surface of the tube
\begin{equation} \label{pressureInt}
    \displaystyle p_{\text{i}} = -\mathbf{T}_{rr}(r_{\text{i}}) =
    \int_{r_{\text{i}}}^{r_{\text{o}}} \frac{\mathbf{T}_{\theta \theta} - \mathbf{T}_{rr}}{r} dr .
\end{equation}
Moreover, we define the axial force $N_{\text{axial}}$ and the reduced axial force $F_{\text{red}}$ by
\begin{equation} \label{AxialForcceN}
 N_{\text{axial}} := 2 \pi \int_{r_{\text{i}}}^{r_{\text{o}}} \mathbf{T}_{zz} \  r  \ dr,
\end{equation}
\begin{equation} \label{ReducedForcceF}
F_{\text{red}}: = N_{\text{axial}} - \pi \ r^2_i \ p_i = \pi \int_{r_{\text{i}}}^{r_{\text{o}}} (2 \mathbf{T}_{zz} - \mathbf{T}_{\theta \theta} -
 \mathbf{T}_{rr}) r  \ dr.
\end{equation}

Due to the incompressibility of the material, the Cauchy stress $\mathbf{T}$ is determined by
the material law uniquely up to a certain hydrostatic component. Note that unknown hydrostatic component becomes
irrelevant when evaluating right-hand sides of \eqref{pressureInt} and \eqref{ReducedForcceF}.
Thus, the internal pressure $p_{\text{i}}$ and the reduced force $F_{\text{red}}$ are unique functions
of $\lambda_{\theta}^{\text{inner}},\lambda_{z}$.
These functions are evaluated numerically in the following way.
The interval $R \in [R_{\text{i}}, R_{\text{o}}]$ is covered by $N$ control points $R_1, R_2, ..., R_N$:
\begin{equation} \label{Subdivision}
dR = \frac{R_{\text{o}} - R_{\text{i}}}{N}, \quad
R_{1} = R_{\text{i}} + \frac{dR}{2}, \quad R_{k} = R_{k-1} + dR \ \text{for} \ k=\overline{2,...,N}.
\end{equation}
These points are needed to track individual particles and they play the same role as the Gauss points in the FEM.
Next, for a given $\lambda^{\text{inner}}_{\theta}$,
 we compute $r_{\text{i}} = \lambda^{\text{inner}}_{\theta} R_{\text{i}}$ and the current coordinates
  of the control points $r_{k} = \sqrt{ \frac{(R_{k}^{2} - R_{\text{i}}^{2})}{\lambda_{z}} - r_{\text{i}}^{2} }, k = \overline{1,...,N}$.
Further, for each control point $r_{k}, k=\overline{1,...,N}$ we compute $\lambda_{\theta}^{k}$ and the deformation
 gradient tensor $\mathbf{F}(\lambda^{k}_{r},\lambda^{k}_{\theta},\lambda^{k}_{z})$,
 substituting $\lambda_{\theta}^{k} = r_k/R_k$, $\lambda_{z}^{k}= \lambda_{z}$, $\lambda^{k}_{r} = 1/(\lambda_{\theta}^{k} \lambda_{z})$ into
 \eqref{DefGradF}.
Using this $\mathbf{F}$, corresponding true stress $\mathbf{T}^{k}$ is computed for $k = \overline{1,...,N}$. Finally, the integrals
 on the right-hand side of \eqref{pressureInt} and \eqref{ReducedForcceF} are approximated by the sum
\begin{equation}\label{pressureSum}
    \displaystyle p_{\text{i}} \approx \sum_{k=1}^{N} \Big(\frac{\mathbf{T}_{\theta \theta}^{k}
    - \mathbf{T}_{rr}^{k}}{r_{k}}\Big) \frac{d r_{k}}{d R_k} \ \Delta R.
\end{equation}
\begin{equation}\label{RedForceSum}
    F_{\text{red}} \approx \pi \sum_{k=1}^{N} \big( 2 \mathbf{T}^{k}_{zz} - \mathbf{T}^{k}_{\theta \theta} -
 \mathbf{T}^{k}_{rr} \big) \frac{d r_{k}}{d R_k} \ \Delta R.
\end{equation}
In this study, all computations are carried out using $N=50$ control points.

In order to simulate the experiment presented in \cite{Wiesemann1995}, we consider the following set up.
The internal pressure is a prescribed function of time: $p_{\text{i}} = p^*_{\text{i}}(t)$. The tube is sealed such that
the internal pressure causes the axial force
$N_{\text{axial}} = \pi \ r^2_i \ p_i$. In terms of the reduced force
this condition reads $F_{\text{red}} = 0$.

The overall loading process is subdivided into $N_{\text{steps}}$ time steps: $t_0 < t_1 <...< t_{N_{\text{steps}}}$.
Consider a generic time step $t_n \mapsto t_{n+1}$.
Let the stretches $\lambda^{\text{inner}}_{\theta}$ and $\lambda_z$ at $t=t_n$
be given by $^{n} \lambda^{\text{inner}}_{\theta}$ and $^{n} \lambda_z$.
To perform the step, the following system of equations is solved with respect to the unknown
$^{n+1} \lambda^{\text{inner}}_{\theta}$ and $^{n+1} \lambda_z$
 \begin{equation}\label{EqOnTimeStep}
 p_{\text{i}}(^{n+1} \lambda^{\text{inner}}_{\theta}, ^{n+1} \lambda_z) = p^*_{\text{i}}(t_{n+1}), \quad
  F_{\text{red}}(^{n+1} \lambda^{\text{inner}}_{\theta}, ^{n+1} \lambda_z) = 0.
\end{equation}
Here, the  dependence of the functions $ p_{\text{i}}(^{n+1} \lambda^{\text{inner}}_{\theta}, ^{n+1} \lambda_z) $ and
$F_{\text{red}}(^{n+1} \lambda^{\text{inner}}_{\theta}, ^{n+1} \lambda_z)$ on the previous history is assumed but omitted for brevity.
Note that any dynamic effects are neglected in this quasi-static problem statement.
In the current study, system \eqref{EqOnTimeStep} is solved numerically using the Newton-Raphson method.

Finally, let us discuss the computation of the stress distribution throughout the tube.
Suppose that $\lambda^{\text{inner}}_{\theta}$ and $\lambda_z$ are known.
The material law allows us to compute the stress tensor $\mathbf{T}^{*k}$
at the control points, which differs from the real true stress
at that point by a certain hydrostatic component, such
that $\mathbf{T}^{k} = \mathbf{T}^{*k} +  p^{k} \mathbf{1}$.
Using the boundary condition, we set $T_{rr}^{N} = 0$. Then, following \eqref{requibl},
the correct values of the radial stresses are computed as
\begin{equation} \label{ruccurence}
T_{rr}^{k} = T_{rr}^{k+1} + \frac{r_{k+1} - r_{k}}{r_{k}}
( T_{rr}^{*k} - T_{\theta \theta}^{*k}), \  k=\overline{N-1,...,1}.
\end{equation}
Basing on the correct value of $T_{rr}^{k}$ we can compute for each control point
the correction term $p^{k} = T_{rr}^{k} - T_{rr}^{*k}$.
Using it, correct stresses in the axial and hoop directions are computed through
$T_{zz}^{k} = T_{zz}^{*k} + p^{k}$ and $T_{\theta \theta}^{k}
= T_{\theta \theta}^{*k} + p^{k}$  for $k = \overline{1,...,N}$.

\section*{Appendix B: Comparison of the semi-analytical procedure with the FEM}

\begin{figure}\centering
    \psfrag{P}[m][][1][0]{$P$, KPa}
    \psfrag{P2}[m][][1][0]{$P$, KPa}
    \psfrag{Lamt}[m][][1][0]{$\lambda_{\theta}$}
    \psfrag{Lamz}[m][][1][0]{$\lambda_{z}$}
    \scalebox{1.0}{
    \includegraphics{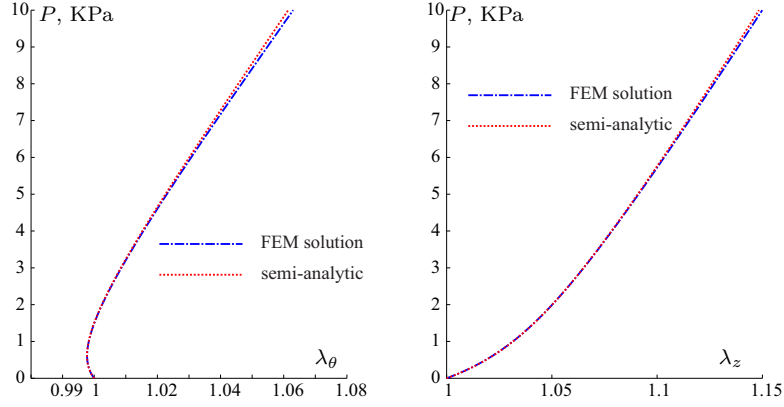}}
    \caption{Comparison of the simulation results obtained by the
    semi-analytical procedure and by the FEM using MSC.MARC. \label{CompFEMAnalyt}}
\end{figure}

The presented composite material model is implemented into the commercial FEM code MSC.MARC using Hypela2 interface.
 The external, middle, and internal
layers of the tube are subdivided into 4, 16, and 10 axisymmetric elements with quadratic approximation of geometry
and displacements. The internal pressure $P$ is applied in 200 steps. Total Lagrange formulation is utilized;
the follower force option is activated. Apart from the applied pressure, axial force $F = \pi \ r^2_i \ P$
is applied to the tube. Material parameters of the composite model from Table \ref{ParamCompositeTube} are used.
As is seen from Figure \ref{CompFEMAnalyt}, the FEM results are in a good agreement
with the simulation results obtained by the semi-analytical method, described in
Appendix A.

\end{document}